\documentclass[aps,10pt,prd,twocolumn,preprintnumbers,amsmath,amssymb,nofootinbib,superscriptaddress,letterpaper,showpacs]{revtex4-1}
\pdfoutput=1
\usepackage[T1]{fontenc}
\usepackage{customcommands}
\usepackage{amsthm}
\usepackage{amsfonts}
\usepackage{xcolor}
\usepackage{slashed}
\usepackage{mathtools}
\usepackage[normalem]{ulem}

\DeclareMathOperator{\arccot}{arccot}

\begin{document}

\title{What spatial geometries do 2+1-dimensional QFT vacua prefer?}

\author{Sebastian Fischetti}
\email{s.fischetti@imperial.ac.uk}

\author{Lucas Wallis}
\email{l.wallis17@imperial.ac.uk}

\author{Toby Wiseman}
\email{t.wiseman@imperial.ac.uk}

\affiliation{Theoretical Physics Group, Blackett Laboratory, Imperial College, London SW7 2AZ, United Kingdom}


\begin{abstract}

We consider relativistic (2+1)-QFTs on a product of time with a two-space and study the vacuum free energy as a functional of the temperature and spatial geometry.  We focus on free scalar and Dirac fields on arbitrary perturbations of flat space, finding that the free energy difference from flat space is finite and always \textit{negative} to leading order in the perturbation.  Thus free (2+1)-QFTs appear to always energetically favor a crumpled space on all scales; at zero temperature this is a purely quantum effect.  Importantly, we show that this quantum effect is non-negligible for the relativistic Dirac degrees of freedom on monolayer graphene even at room temperature, so we argue that this vacuum energy effect should be included for a proper analysis of the equilibrium configuration of graphene or similar materials.

\end{abstract} 

\maketitle


\section{Introduction}

The presence of matter gives a surface embedded in an ambient space an energy.  This matter  may be external to the surface -- like the pressure of air on a soap bubble -- or may comprise the material nature of the surface itself -- like a membrane with surface tension and bending energy.  These energies determine the equilibrium (i.e.~static) configuration of such a surface: for instance, the presence of surface tension tends to make membranes favor (smooth) minimal-area configurations, while finite-temperature thermodynamic effects may render membranes unstable to crumpling or rippling~\cite{Order,MembraneBook,MembraneReview}. 

In this Letter we initiate a study of the free energy contribution to the equilibrium configuration of a surface due to free relativistic quantized matter fields living on it.  In particular, we include zero-temperature (Casimir) effects.  Such relativistic quantum fields occur in various physical settings: for example, in graphene and related materials, the electronic structure gives rise to an effective description in terms of relativistic Dirac fermions propagating on the two-dimensional crystal \cite{Graphene,GrapheneDirac1,GrapheneDirac2}.  In cosmology domain wall defects may exist \cite{CosmoBook} and could carry upon them relativistic degrees of freedom.  More exotically, in braneworld models our universe is itself a surface on which the Standard Model fields live \cite{ArkaniHamed:1998rs,Randall:1999ee}.

The setting is then (2+1)-d QFT on a product of time with a two-space.  By studying both the free non-minimally coupled scalar\footnote{Note that free massless vector fields are equivalent to free minimally coupled massless scalar fields by duality in~$(2+1)$-d, so our analysis indirectly includes massless vector fields as well.} and the free Dirac fermion we will see that such a field \textit{lowers} the free energy of the surface on which it lives when the surface is deformed away from being intrinsically flat\footnote{Our analysis was motivated by holographic considerations: for~$(2+1)$-d conformal field theories (CFTs), flat space is energetically disfavored at zero temperature.  This was shown globally in holographic CFTs and perturbatively in general~\cite{FisHic16,FisWis17}.}.  This energy difference is UV finite (and thus well-defined) and present at any temperature including~$T = 0$ (in which case it can be interpreted as a Casimir effect) both for massless and massive fields, and any scalar non-minimal coupling.  It is then natural to wonder whether a classical membrane action is able to counteract this quantum tendency to crumple.  We will perform a na\"ive analysis of this question for monolayer graphene, which is indeed seen to ripple on short scales \cite{GrapheneRippleExpt,GrapheneRippleMC}. We show that at room temperature the \emph{quantum} vacuum energy of the  Dirac fermions give a scale at which one would expect crumpling effects on the order of the lattice spacing.  The effective membrane description which would validate our analysis breaks down at this scale, so our results make no definitive statement about the rippling of graphene.  However, they do indicate that a careful consideration of these quantum effects is important for a proper treatment of equilibrium configurations of graphene and similar materials even at room temperature.

\section{Free Energy Difference}

We consider a spacetime which is a product of time with a two-space~$\Sigma$ (for now taken to be general).  Since we are interested in QFT at finite temperature~$T$ we work in Euclidean time, so the metric is\footnote{Unless otherwise stated we use natural units $\hbar = c = k_B = 1$ with~$c$ the ``effective'' speed of light of the relativistic fields (not necessarily equal to the actual speed of light).}
\be
\label{eq:metric}
ds^2 = d\tau^2 + ds^2_\Sigma
\ee
with~$\tau$ periodic with period~$\beta = 1/T$.  We will consider a free scalar~$\phi$ and Dirac spinor~$\psi$ with equations of motion
\be
\label{eq:EOM}
\left(-\nabla^2 + \xi R + M^2 \right) \phi = 0 \, , \quad \left(\slashed{D} + M \right) \psi = 0
\ee
respectively, where~$R$ is the Ricci scalar and~$\slashed{D}$ is understood as being defined by the spin connection (our conventions can be found in the Supplemental Material).

The free energy~$F[\Sigma]$ is a functional of the geometry~$\Sigma$ (and temperature~$T$) and is given in terms of the partition function~$Z[\Sigma]$ as~$F = - T \ln Z$.  We are specifically interested in the difference~$\Delta F$ between the free energy on~$\Sigma$ and some reference background space~$\overline{\Sigma}$ at the same temperature, which satisfies
\be
\label{eq:DeltaFdef}
e^{-\beta \Delta F} = \frac{Z[\Sigma]}{Z[\overline{\Sigma}]} = \frac{\int \Dcal \Phi \, e^{-S_\Sigma[\Phi]}}{\int \Dcal \Phi \, e^{-S_{\overline{\Sigma}}[\Phi]}} = \left\langle e^{-\Delta S} \right\rangle_{\overline{\Sigma}},
\ee
where~$\Phi$ stands for the matter field (scalar or fermion) being integrated over in the path integral,~$\Delta S = S_\Sigma - S_{\overline{\Sigma}}$ is the difference between the action on~$S^1 \times \Sigma$ and~$S^1 \times \overline{\Sigma}$, and the expectation value is defined by the path integral on the background geometry~$S^1 \times \overline{\Sigma}$.  To evaluate~$\Delta F$, recall that for free fields the path integrals in~\eqref{eq:DeltaFdef} yield functional determinants, giving
\be
\label{eq:L}
Z = (\det \Lcal)^q \mbox{ with } \Lcal = -\partial_\tau^2 + \Ocal + M^2,
\ee
where~$q = -1/2$ ($+1$) for the scalar (fermion),~$\Ocal$ is an elliptic self-adjoint scalar operator on~$\Sigma$ given explicitly in~\eqref{eqs:Oops} below, and the  determinant is evaluated over Matsubara frequencies on the thermal circle (with appropriate periodicity or antiperiodicity in the scalar and fermion cases respectively).  For the scalar,~\eqref{eq:L} is obtained straightforwardly. The fermion case is more subtle, and we leave full details to the Supplemental Material.  In short, a direct path integral yields~$Z = \det(i\slashed{D} - iM)$.  However, by exploiting the direct product structure of the metric~\eqref{eq:metric} along with the fact that the two-dimensional rotation group only has a single generator, we 
may eliminate the spinor structure and reduce the determinant to that of an elliptic operator of the form~\eqref{eq:L} with the determinant taken over the space of complex functions with  antiperiodicity on the thermal circle.  

The free energy can then be evaluated via heat kernel methods~\cite{Vas03}: defining the heat kernel as~$K_\Lcal(t) \equiv \Tr(e^{-t\Lcal}) = \sum_i e^{-t \lambda_i}$ (with~$\lambda_i$ the eigenvalues of~$\Lcal$), one obtains
\be
\beta \Delta F = q \int_0^\infty \frac{dt}{t} \Delta K_\Lcal(t),
\ee
where~$\Delta K_\Lcal(t) \equiv K_\Lcal(t) - K_{\overline{\Lcal}}(t)$.  This expression is UV divergent unless~$\Delta K_\Lcal(t)$ vanishes at~$t = 0$; this condition can be ensured by an appropriate choice of background~$\overline{\Sigma}$.  Specifically, the heat kernel expansion gives~\cite{Vas03}
\be
\label{eq:heatkernelexpansion}
K_\Lcal(t) = \beta \left[\frac{c_1 V_\Sigma}{t^{3/2}} + \frac{c_2 \chi_\Sigma + c_3 V_\Sigma M^2}{t^{1/2}} + O(t^{1/2}) \right],
\ee
where~$V_\Sigma$ and~$\chi_\Sigma$ are the volume\footnote{Suitably IR regulated if~$\Sigma$ is non-compact.} and Euler number of~$\Sigma$, respectively, and~$c_1$,~$c_2$, and~$c_3$ are dimensionless constants independent of the geometry (though they depend on the choice of matter field).  Thus requiring that~$\Delta F$ be UV-finite 
only
imposes that we choose a background geometry~$\overline{\Sigma}$ with the same volume and topology as~$\Sigma$.
It is worth emphasizing that although the undifferenced functional determinant~$\det\Lcal$ is UV-divergent, we do not need to invoke any regularization to evaluate the \textit{differenced} free energy\footnote{Note that since~$\det \Lcal$ is UV-divergent,~$\Delta F$ is not necessarily the same as a difference of separately renormalized free energies on~$\Sigma$ and~$\overline{\Sigma}$, which could contain renormalization ambiguities and therefore be unphysical.}.  It is also worth noting that in higher dimensions, the expansion~\eqref{eq:heatkernelexpansion} contains non-topological curvature invariants of~$\Sigma$; thus obtaining a UV-finite free energy difference would require a careful matching of these invariants on~$\Sigma$ and~$\overline{\Sigma}$ (in contrast with the heuristic expectation that ``energy differences are always UV-finite'').

Now we specialize to our case of interest.  Ultimately we wish to take~$\Sigma$ to be a deformation of flat space, $\overline{\Sigma}$.  Since these are two-dimensional we introduce conformally flat coordinates~$x^A$, in terms of which the metrics on~$\Sigma$ and~$\overline{\Sigma}$ take the form
\be
\label{eq:sigmametric}
ds^2_\Sigma = e^{2 f(x)} \delta_{AB} dx^A \, dx^B \, , \quad ds^2_{\overline{\Sigma}} =  \delta_{AB} dx^A \, dx^B.
\ee
In order to have good control over the spectrum of~$\Lcal$ (which is essential for computing the heat kernel), we compactify these to tori~$\Sigma_L$,~$\overline{\Sigma}_L$ via the identifications~$x^A \sim x^A + L_A$ with $L_1 = L$ and $L_2 = r L$. We consider a family of deformations~$f_L(x)$ so that as $L \to \infty$ (with $r$ fixed) we recover~\eqref{eq:sigmametric} with the~$x^A$ uncompactified.  Moreover, at any finite~$L$, we may choose~$f_L$ such that~$V_{\Sigma_L} = V_{\overline{\Sigma}_L}$.  By the arguments above, this condition will ensure that for every~$L$, the free energy difference between the deformed and flat torus will be UV finite.\footnote{For finite~$L$ one may choose between periodic and antiperiodic boundary conditions for the fermion on the torus cycles; since our torus is only an IR regulator, and this distinction vanishes in the limit~$L \to \infty$, we take the periodic case for simplicity.}

Our object of interest is the free energy difference with this IR regulator removed:
\be
\beta \Delta F_\infty[f] \equiv q \lim_{L \to \infty} \int_0^\infty \frac{dt}{t} \Delta K_\Lcal[f_L;L](t),
\ee
with~$\Delta K_\Lcal[f_L;L](t) \equiv K_\Lcal[f_L;L](t) - K_\Lcal[0;L](t)$.  For notational convenience we will henceforth forego writing the arguments of these functionals explicitly, and we will drop the subscripts~$L$ on~$f_L$ and~$\infty$ on~$\Delta F_\infty$.  Using~\eqref{eq:L}, we finally obtain
\be
\label{eq:diffF}
\beta \Delta F[f] = q \lim_{L \to \infty}  \int_{0}^\infty \frac{dt}{t} e^{- M^2 t} \Theta( T^2 t) \Delta K_\Ocal(t),
\ee
where the sum over Matsubara frequencies yields
\be
\label{eq:Thetadef}
\Theta(\zeta) = \sum_{n=-\infty}^{\infty} e^{- (2 \pi)^2 (n-q+1/2)^2 \zeta}
\ee
and the operators~$\Ocal$ are given explicitly in terms of~$f$ as
\begin{subequations}
\label{eqs:Oops}
\begin{align}
\label{eq:opscalar}
\Ocal_\mathrm{s} &= - e^{-2 f} \left( \overline{\grad}^2 + 2 \xi (\overline{\grad}^2 f) \right), \\
\label{eq:opdirac}
\Ocal_\mathrm{f} &= \Ocal_\mathrm{s}|_{\xi = 1/4} - e^{-2 f} \left( i \, \bar{\star} \left( df \wedge d \right)  - \frac{( \overline{\grad}_A f)^2}{4} \right),
\end{align}
\end{subequations}
with~$\overline{\grad}_A$ and~$\bar{\star}$ the covariant derivative and Hodge dual on the flat background~$f = 0$, and the subscripts~$s$ and~$f$ denoting the scalar and fermion.

\section{Perturbation Theory}

We introduce a perturbation parameter~$\eps$ to expand our deformation $\Sigma$ of the flat~$\overline{\Sigma}$ as
\be
f = \eps f^{(1)} + \eps^2 f^{(2)} + O(\eps^3).
\ee  
Preservation of the volume requires
\be
\label{eq:volcondition}
\int d^2 x \, f^{(1)} = 0, \qquad \int d^2 x \left(f^{(2)} + \left(f^{(1)}\right)^2 \right) = 0.
\ee
We denote by~$\lambda_i$ and~$h_i(x)$ the eigenvalues and eigenfunctions of~$\Ocal$, so in order to compute~$\Delta K_\Ocal(t)$ perturbatively we must compute the perturbative corrections to~$\lambda_i$.  We expand~$\Ocal$ as
\be
\Ocal = -\overline{\grad}^2 + \eps \Ocal^{(1)} + \eps^2 \Ocal^{(2)} + O(\eps^3);
\ee
explicit expressions for~$\Ocal^{(1)}$ and~$\Ocal^{(2)}$ can be obtained by expanding the form of~$\Ocal$ given in~\eqref{eqs:Oops}.  Continuing to use bars to denote unperturbed objects, we likewise expand its eigenvalues and eigenfunctions as
\bea
\lambda_i &= \bar{\lambda}_i + \eps \lambda_i^{(1)} + \eps^2 \lambda_i^{(2)} + O(\eps^3), \\
h_i &= \bar{h}_i + \eps \sum_{j} b_{ij}^{(1)} \bar{h}_j + \eps^2 \sum_{j} b_{ij}^{(2)} \bar{h}_j + O(\eps^3) \label{eq:hexpansion},
\eea
so the~$\bar{h}_i$ are eigenfunctions of the flat space Laplacian with corresponding eigenvalues~$\bar{\lambda}_i$.  We choose these eigenfunctions to be normalized as
\be
\langle\bar{h}_i|\bar{h}_j\rangle \equiv \int d^2 x \, \bar{h}_i^* \bar{h}_j = \delta_{ij}.
\ee 
Then defining
\bea
P^{(1)}_{ij} &= \langle \bar{h}_i | \Ocal^{(1)} \bar{h}_j \rangle, \\
P^{(2)}_{ij} &= \langle \bar{h}_i | \Ocal^{(2)} \bar{h}_j \rangle + \sum_{k; \, \bar{\lambda}_k \neq \bar{\lambda}_i} \frac{P^{(1)}_{ik} P^{(1)}_{kj}}{\bar{\lambda}_i - \bar{\lambda}_k},
\eea
standard perturbation theory yields the eigenvalue shifts
\be
\label{eq:perteigvals}
\lambda_i^{(1)} =  P^{(1)}_{ii}, \quad \lambda_i^{(2)} =  P^{(2)}_{ii} \quad \mbox{(no sum)}.
\ee

Note that we have glossed over a subtlety: recall from QM perturbation theory that the presence of degenerate subspaces imposes additional constraints on the unperturbed eigenfunctions~$\bar{h}_i$ for the expansion~\eqref{eq:hexpansion} to be consistent.  The first order eigenvalue problem requires we arrange our basis~$\bar{h}_i$ such that~$P^{(1)}_{ij}$ is diagonal within such subspaces (i.e.~if~$\bar{\lambda}_i = \bar{\lambda}_j$ but~$i \neq j$, then~$P^{(1)}_{ij} = 0$).  If any degeneracies remain at first order, we must further ensure at second order that~$P^{(2)}_{ij}$ be diagonal in the remaining degenerate subspaces. We discuss this issue explicitly in the Supplemental Material.

Finally, we expand the heat kernel as
\be
\Delta K_\Ocal(t) = \eps K^{(1)}(t) + \eps^2 K^{(2)}(t) + O(\eps^3),
\ee
with
\bea
K^{(1)}(t) &= -t \sum_i e^{- \bar{\lambda}_i t} P^{(1)}_{ii}, \\
K^{(2)}(t) &= t \sum_i e^{- \bar{\lambda}_i t} \left( \frac{t}{2} \left(P^{(1)}_{ii}\right)^2  - P^{(2)}_{ii} \right) \label{eq:K2finiteL}
\eea

\section{Results}

In order to perform our computations we Fourier decompose the perturbation
\bea
f^{(1)}(x) &= \frac{(2\pi)^2}{r L^2} \sum_{\vec{N}} \tilde{f}_{\vec{N}}^{(1)} e^{2 \pi i \left(n_1 x^1 + n_2 x^2/r \right)/L}, \\
	 &\to \int d^2 k \, \tilde{f}^{(1)}(\vec{k}) e^{i \vec{k} \cdot \vec{x}} \mbox{ as } L \to \infty,
\eea
where the sum runs over all pairs of integers~$\vec{N} = \{n_1,n_2\}$, and the second line defines~$k_A = \lim_{L \to \infty} 2\pi n_A/L_A$.

An explicit calculation on the torus for fixed~$L$ reveals that (for both the scalar and fermion) while the eigenvalues are indeed shifted at first order in~$\eps$, their contribution to the heat kernel vanishes:~$K^{(1)} = 0$.  The leading order perturbation to the heat kernel is then the second order term~$K^{(2)}$.  A lengthy but straightforward computation yields the finite-$L$ expressions presented in the Supplemental Material; in the limit~$L \to \infty$ they become
\be
\label{eq:heatkernelexpr}
K^{(2)}(t) = t \int d^2k \, k^4 \left| \tilde{f}^{(1)}(\vec{k}) \right|^2 I(k^2 t)
\ee
with~$k = |\vec{k}|$,
\begin{subequations}
\begin{multline}
\label{eq:scalar}
I_\mathrm{s}(\zeta) = -\frac{\pi}{4 \zeta^2} \left[6 +  \zeta ( 1 - 8 \xi) \phantom{ \frac{\sqrt{\zeta}}{ 2} } \right. \\
		\left. - \left( 6 + 2 \zeta (1 - 4 \xi) + \frac{\zeta^2}{2}  (1 - 4 \xi)^2 \right) \Fcal\left( \frac{\sqrt{\zeta}}{ 2} \right) \right],
\end{multline}
\be
\label{eq:dirac}
I_\mathrm{f}(\zeta) = \frac{\pi}{4 \zeta^2} \left[\left(6 + \zeta\right) \Fcal\left(\frac{\sqrt{\zeta}}{ 2} \right) - 6 \right],
\ee
\end{subequations}
and~$\Fcal(\zeta) = \zeta^{-1} e^{-\zeta^2} \int_0^{\zeta} d\zeta' \, e^{(\zeta')^2}$.  Thus using~\eqref{eq:diffF} we find
\be
\label{eq:diffFexpr}
\Delta F = -\eps^2 \int d^2k \, a(k) \left| \tilde{f}^{(1)}(\vec{k}) \right|^2,
\ee
with
\be
\label{eq:adef}
a(k) \equiv -q T k^4 \int_0^\infty dt \, e^{-M^2 t} \Theta(T^2 t) I(k^2 t).
\ee
A few comments are in order.  Firstly we see the leading variation in~$\Delta F$ is quadratic in~$\eps$.  
Next, as $L \to \infty$ the volume constraint~$V_{\Sigma_L} = V_{\overline{\Sigma}_L}$ becomes the condition that the variation of the volume $\int d^2x \sqrt{g}$ vanishes; for $f^{(1)}$ this simply imposes no constant Fourier component.
We have that~$I(\zeta)$ is finite and~$\Theta(\zeta)$ is~$O(\zeta^{-1/2})$ at small~$\zeta$, and thus~$\Delta F$ is UV-finite.  Likewise, since~$I(\zeta)$ and~$\Theta(\zeta)$ are finite at large~$\zeta$,~$\Delta F$ is also IR-finite for~$M > 0$; in fact, the large-$\zeta$ decay of~$I(\zeta)$ also implies IR finiteness in the massless case $M = 0$ for both the fermion and minimally-coupled scalar~($\xi = 0$)\footnote{For the massless scalar with non-minimal coupling~$\xi \neq 0$,~$\Delta F_\mathrm{s}$ is  IR divergent since the flat space zero eigenvalue aquires a negative contribution due to the scalar curvature coupling.  This is reflected in the~$\ln(\ell_M/\ell)$ corrections mentioned in Table~\ref{tab:scalings}.}.  Finally, a key physical point is that, as can be seen by explicitly plotting\footnote{It is possible to prove that~$I_\mathrm{f} < 0$ without resorting to plotting it; we have not been able to find as elegant of a proof for~$I_\mathrm{s}$.},~$q I(\zeta) < 0$ for all~$\zeta > 0$ (and all~$\xi$ for the scalar), implying that for any (non-constant)~$f$ the free energy difference is strictly negative to leading order in~$\eps$: $\Delta F < 0$.

The form of the expression~\eqref{eq:adef}, along with the asymptotic behaviors of~$\Theta(\zeta)$ and~$I(\zeta)$, makes it possible to derive scaling relations.  Specifically, defining~$\ell_M = \hbar/(c M)$ to be the (reduced) Compton wavelength,~$\ell_T = \hbar c/(k_B T)$ to be a thermal wavelength, and~$\ell$ to be the characteristic length scale of~$f$,~$\Delta F$ scales as in Table~\ref{tab:scalings}.  Thus at small temperatures -- in which~$\Delta F$ becomes the energy difference~$\Delta E$ -- the effect is a purely quantum one:~$\Delta E \sim -\eps^2 \hbar c/\ell$ for~$\ell \ll \ell_M$.
We are exploring numerical methods to verify these perturbative results, and also understand the non-perturbative regime \cite{ongoing}.

\begin{table}
\begin{tabular}{|c|c|c|}
\hline
 & $-\Delta F_\mathrm{s}/(\eps^2 \hbar c/\ell)$ & $-\Delta F_\mathrm{f}/(\eps^2 \hbar c/\ell)$ \\
\hline
$\ell_T \gg \ell \gg \ell_M$ & $\ell_M/\ell$ & $\ell_M/\ell$ \\
$\ell_T \gg \ell_M \gg \ell$ & $1$ & $1$ \\
$\ell \gg \ell_T \gg \ell_M$ & $\ell_M/\ell$ & $\ell_M/\ell$ \\
$\ell \gg \ell_M \gg \ell_T$ & $\ell_M^2/(\ell\ell_T)$ & $\ell_T/\ell$ \\
$\ell_M \gg \ell_T \gg \ell$ & $1$ & $1$ \\
$\ell_M \gg \ell \gg \ell_T$ & $\ell/\ell_T$ & $\ell_T/\ell$ \\
\hline
\end{tabular}
\caption{The scaling of~$\Delta F$ for the minimally coupled free scalar field and Dirac fermion for different relative magnitudes of~$\ell$,~$\ell_M$, and~$\ell_T$.  Note that for the non-minimally coupled scalar (i.e.~$\xi \neq 0$), factors of~$\ln(\ell_M/\ell)$ appear in the last two rows.}
\label{tab:scalings}
\end{table}

As a final note, the small-temperature limit~$\ell_T \gg \max[\ell,\ell_M]$ is in fact analytically tractable: Poisson resummation gives~$\Theta(T^2 t) = \beta/\sqrt{4 \pi t}$ up to terms that are exponentially suppressed in~$\beta^2/t$, which allows us to compute~$a(k)$ explicitly as\footnote{In the massless limit these agree precisely with the energy in~\cite{FisWis17} for the massless scalar CFT ($\xi=1/8$) and free fermion CFT (with their appropriate central charges $c_T = (3/2)/(4\pi)^2$ and $3/(4\pi)^2$ respectively).}
\begin{widetext}
\bea
\label{eq:lowTscalar}
a^{(T = 0)}_\mathrm{s}(k)& = \frac{\pi k^3}{128} \left[\frac{2(3-32\xi)M}{k} - \frac{24M^3}{k^3} + \left(3-32\xi + 128\xi^2 - 8(1-16\xi)\frac{M^2}{k^2} + \frac{48 M^4}{k^4} \right) \arccot\left( \frac{2 M}{k} \right) \right], \\
\label{eq:lowTfermion}
a^{(T = 0)}_\mathrm{f}(k) &= \frac{\pi k^3}{64} \left[\frac{2M}{k} + \frac{24M^3}{k^3}  + \left(1 - \frac{8 M^2}{k^2} - \frac{48 M^4}{k^4} \right) \arccot\left( \frac{2 M}{k} \right) \right].
\eea
\end{widetext}

\section{Membrane Crumpling}

We have seen that free relativistic (2+1)-dimensional degrees of freedom on deformations of flat space 
that have UV finite free energy difference from flat space always energetically prefer the deformation, for any temperature.
Let us now consider how this effect competes with a membrane's bending energy (which at zero temperature favors a flat geometry) 
if it carries such degrees of freedom.

Consider three-dimensional flat space with Cartesian coordinates $\{X^A, Z\}$ and parametrize a surface in it by $X^A = x^A + \eps v^A(x^B)$, $Z = \sqrt{\eps} \, h(x^A)$. Then for small~$\eps$ and suitable~$v^A$, the intrinsic metric on the membrane in the coordinates~$x^A$ is as in equation~\eqref{eq:sigmametric} with~$-\overline{\grad}^2 f = \eps \det( \partial_A \partial_B h )$\footnote{
The constant part of $f$ is not determined by this relation, thus  we may choose it so that $f^{(1)}$ has no constant Fourier component. 
}. 
The bending energy due to extrinsic curvature is
\be
H = \eps \kappa \int d^2x (\overline{\grad}^2 h)^2,
\ee
where~$\kappa$ is the bending rigidity.  If the membrane is deformed from flat over a region of characteristic size $\ell \ll \ell_M$, then the (positive) bending energy~$E_B$ and (negative) vacuum energy~$E_Q$ (at zero temperature) for $N$ free relativistic quantum fields are parametrically given as
\be
\label{eq:diffFmembrane}
E_{B} \sim \eps \kappa, \qquad E_{Q} \sim -\eps^2 N \frac{\hbar c}{\ell}.
\ee
The ground state equilibrium configuration of the membrane should minimize~$E = E_B + E_Q$.  One might expect that because~$E_B$ is lower order in~$\eps$ than~$E_Q$, a perturbative analysis guarantees that~$E > 0$ for any deformation of flat space.  However, note that~$E_B$ and~$E_Q$ have different scale dependence, with~$E_Q$ dominating at sufficiently small scales.  
Defining $\ell_\mathrm{crumple} \equiv N \hbar c/ \kappa$ and
%
noting that $\epsilon$ and $\ell_\mathrm{crumple} / \ell$ are independent,  if
%
~$\ell / \ell_\mathrm{crumple} \lesssim \epsilon \ll 1$ then~$E_Q$ can be comparable to and even dominate~$E_B$ while still being in the perturbative regime.  Whether or not~$E$ actually decreases for (sufficiently large) deformations of flat space -- therefore implying that the membrane's equilibrium configuration is crumpled at a sufficiently small scale relative to $\ell_\mathrm{crumple}$ -- then depends on nonlinear and higher-derivative contributions to its bending action and whether or not these are relevant at
scales up to~$\ell_\mathrm{crumple}$ at amplitudes~$O(\eps^2)$.
Hence $\ell_\mathrm{crumple}$ defines a scale below which a membrane has the potential to crumple.

It is instructive to consider the case of a graphene monolayer, for which the bending rigidity is~$\kappa \sim 1$~eV, the unit cell has size~$\ell_\mathrm{cell} \sim 1$~\AA, and the relativistic fields are two Dirac fermions with effective speed $c \sim c_\mathrm{light}/100$, with $c_\mathrm{light}$ the actual speed of light \cite{GrapheneDirac1,GrapheneRippleMC,GrapheneDirac,Fialkovsky:2016kio}.  Our effective membrane description is valid for~$\ell \gg \ell_\mathrm{cell}$, while from Table~\ref{tab:scalings} the scaling properties~\eqref{eq:diffFmembrane} are valid at room temperature for~$\ell \ll \ell_{T = 300 \, K} \sim 10^3 \ell_\mathrm{cell}$.  Computing the 
potential 
crumpling scale, we find~$\ell_\mathrm{crumple} \sim 10 \ell_\mathrm{cell}$, which is sufficiently close to~$\ell_\mathrm{cell}$ to make our effective membrane description suspect.  Thus while this na\"ive analysis is insufficient to imply the existence of a crumpled equilibrium configuration for graphene, it \textit{does} indicate that 
\emph{long range quantum properties}
 of the conduction electrons (which give rise to the effective Dirac fermions) are important for understanding the energetics of equilibrium monolayer graphene even at room temperature; such effects are presumably highly challenging to correctly incorporate into Monte Carlo or \emph{ab initio} quantum simulations.
 Indeed, it is intriguing to note that for freely suspended graphene at room temperature, one does see low amplitude ripples on short scales $\sim 50$~\AA, close to our~$\ell_\mathrm{crumple}$ \cite{GrapheneRippleExpt}.

We emphasize that in future two-dimensional crystal materials whose electronic structures similarly give rise to Dirac fermions (or perhaps scalars or vectors), the situation may be different.  In particular, if one wishes to have such a monolayer material that is flat on scales above the unit cell scale~$\ell_\mathrm{cell}$, one presumably requires $N \hbar c / \kappa \lesssim \ell_\mathrm{cell}$, which may be regarded as a bound on the speed or number of relativistic species, given the bending mechanics of the crystal.


\section*{Acknowledgements}

This work was supported by the STFC grants ST/L00044X/1 and ST/P000762/1.


\bibliographystyle{apsrev4-1}
\bibliography{spectrum}

\appendix

\section*{Supplemental Material}

\subsection{Fermion Partition Function}

We will follow the Clifford algebra conventions of~\cite{FreVan}: in Euclidean signature, the Clifford algebra is
\be
\{ \gamma^\mu, \gamma^\nu \} = 2\delta^{\mu\nu},
\ee
which allows us to take the~$\gamma^\mu$ to be Hermitian.  With such conventions, a natural choice of representation of the gamma matrices in three dimensions is~$\gamma^\mu = \sigma^\mu$ with~$\sigma^\mu$ the Pauli matrices, though we note that none of our statements will depend on such a choice.  Scalars are formed from spinors~$\chi$,~$\psi$ as the bilinears~$\bar{\chi} \psi$ with~$\bar{\chi} = \chi^\dag$, and the massive Euclidean Dirac action on a curved space with metric~$g_{ab}$ is
\be
S_E[\bar{\psi},\psi] = \int d^3x \sqrt{g} \, \bar{\psi}(i \slashed{D} - iM) \psi,
\ee
where~$\slashed{D} = \gamma^\mu (e_\mu)^a D_a$ with~$\{(e_\mu)^a\}$ for~$\mu = 1$, 2, 3 a vielbein,
\be
D_a = \grad_a + \frac{1}{2} \omega_{a\mu\nu} S^{\mu\nu},
\ee
$\grad_a$ the usual Riemann connection compatible with~$g_{ab}$,~$S^{\mu\nu} = [\gamma^\mu,\gamma^\nu]/4$ the generators of the Lorentz group, and~$\omega_{a\mu\nu} = (e_\mu)^b \grad_a (e_\nu)_b$ the spin connection.  Note that the operator~$i\slashed{D}$ is self-adjoint, but the~$i$ in the mass term renders the Euclidean action non-Hermitian.  This factor of~$i$ is necessary to ensure that the action obeys the Osterwalder-Schrader positivity conditions; see e.g.~\cite{Wet10} for a discussion of such subtleties associated with spinors in Euclidean space.

Performing the path integral yields
\be
Z = \int \Dcal \bar{\psi} \, \Dcal \psi \, e^{-S_E[\bar{\psi},\psi]} = \det(i\slashed{D} - iM).
\ee
Because~$i\slashed{D}$ is self-adjoint, its eigenvalues are real.  Moreover, in the direct product geometry~\eqref{eq:metric}, eigenspinors of~$i\slashed{D}$ can be decomposed into Fourier modes~$\psi = e^{-i\Omega_n \tau} \psi_\Sigma$, with~$\psi_\Sigma$ a spinor on~$\Sigma$ and~$\Omega_n = (2n+1)\pi/\beta$ a Matsubara frequency (with~$n \in \mathbb{Z}$).  It is then straightforward to show that if~$e^{-i\Omega_n \tau} \psi_\Sigma$ is an eigenspinor of~$i\slashed{D}$ with eigenvalue~$\lambda$, then~$e^{i\Omega_n \tau} \gamma^\tau \psi_\Sigma$ is an eigenspinor with eigenvalue~$-\lambda$.  Thus the spectrum of~$i\slashed{D}$ on the background~\eqref{eq:metric} is symmetric about zero\footnote{Note that the direct product structure of~\eqref{eq:metric} was crucial; in a general odd-dimensional geometry the spectrum of~$i\slashed{D}$ need not be symmetric~\cite{DowLev13}.}, so we have
\be
Z^2 = \det(i\slashed{D} - iM) \det(-i\slashed{D} - iM) = \det(\slashed{D}^2 - M^2).
\ee
(See e.g.~\cite{DeB01} for more on this trick in~$d = 4$.)  Now, writing the metric on~$\Sigma$ in the conformally flat form~\eqref{eq:sigmametric}, one can evaluate~$\slashed{D}^2 - M^2$.  Noting that there is only one generator~$S^{12} = (i/2)\gamma^\tau$ of rotations in two dimensions, we obtain
\be
\slashed{D}^2 - M^2 = -\Lcal P_L - \Lcal^* P_R,
\ee
where~$P_{L,R} = (1 \pm \gamma^\tau)/2$ are left and right Weyl projectors on~$\Sigma$ and~$\Lcal$ is as given in~\eqref{eq:L}.  Decomposing~$\psi = e^{-i\Omega_n \tau} \psi_\Sigma$, we see that~$\Lcal$ and~$\Lcal^*$ act only on left- and right-helicity Weyl spinors~$P_L \psi_\Sigma$,~$P_R \psi_\Sigma$, respectively.  Since these spinors only have one component each, we may just interpret~$\Lcal$ and~$\Lcal^*$ as acting only on complex functions (albeit with antiperiodic boundary conditions on the thermal circle).  We therefore have
\be
\det(-\Lcal P_L - \Lcal^* P_R) = \det(-\Lcal) \det(-\Lcal^*) = (\det \Lcal)^2,
\ee
where in the second expression we take the determinants only over the space of functions on which~$\Lcal$ and~$\Lcal^*$ act, and in the last equality we noted that because~$\Lcal$ is self-adjoint (with respect to the usual~$L_2$ norm),~$\Lcal$ and~$\Lcal^*$ have the same spectrum (and thus determinant).  Thus the partition function for the fermion can be evaluated by just taking the functional determinant of a scalar differential operator acting on complex functions.
\\

\subsection{Finite-$L$ Heat Kernels}

Here we provide more details on the computation of the heat kernel at finite~$L$.  First, in order to deal with the issue of eigenfunction degeneracy, it is convenient to take~$r^2$ irrational (so that no eigenvalue~$\bar{\lambda}_i$ has degeneracy greater than four) and choose the label~$i$ to consist of~$\{\vec{N}^+,\vec{S}\}$, where~$\vec{N}^+ = \{n_1^+,n_2^+\}$ is a pair of nonnegative integers and~$\vec{S} = \{s_1,s_2\}$ is a pair of signs, with~$s_A = \pm 1$ if~$n_A^+ \neq 0$ and~$s_A = 0$ if~$n_A^+ = 0$.  The values of~$\vec{S}$ index the degenerate subspaces; for a given~$\vec{N}^+$, there are~$d_{\vec{N}^+} = (2-\delta_{n_1^+,0})(2-\delta_{n_2^+,0})$ possible such values.  The eigenvalues of~$-\overline{\grad}^2$ are then given by~$\vec{N}^+$ as~$\bar{\lambda}_{\vec{N}^+} = (2\pi/L)^2((n_1^+)^2 + (n_2^+)^2/r^2)$ and have degeneracy~$d_{\vec{N}^+}$, while we write the eigenfunctions as
\be
\bar{h}_{\vec{N}^+,\vec{S}}(x) = \frac{1}{\sqrt{r}\, L} \sum_{\vec{S}'} c^{(\vec{N}^+)}_{\vec{S} \vec{S'}} e^{2 \pi i \sum_{A = 1}^2 s'_A n_A^+ x^A/L_A},
\ee
where the sum runs over all~$d_{\vec{N}^+}$ possible choices of~$\vec{S'}$ and for fixed~$\vec{N}^+$,~$c^{(\vec{N}^+)}_{\vec{S}\vec{S}'}$ is an arbitrary~$d_{\vec{N}^+} \times d_{\vec{N}^+}$ unitary matrix.  In other words, for given~$\vec{N}^+$ the~$\bar{h}_{\vec{N}^+,\vec{S}}$ form an arbitrary orthonormal basis of the degeneracy subspace of~$-\overline{\grad}^2$ with eigenvalue~$\bar{\lambda}_{\vec{N}^+}$; the freedom to choose this basis is what allows us to satisfy the perturbation theory constraints on~$\bar{h}_{\vec{N}^+,\vec{S}}$.

Using this formalism, we may compute the second-order correction to the heat kernel at finite~$L$ using~\eqref{eq:K2finiteL}.  After some rearrangement, we find
\begin{widetext}
\begin{subequations}
\begin{multline}
K^{(2)}_\mathrm{s}(t) = t \frac{4(2 \pi)^4}{(L_1 L_2)^2} \left[ \frac{t}{2} \sum_{\vec{N}^+}  e^{-\bar{\lambda}_{\vec{N}^+} t} \sum_{\mathclap{\vec{S},\vec{S}', \vec{S}' \ne \vec{S}}} \left| \tilde{f}^{(1)}_{\Delta \vec{S} \vec{N}^+} \right|^2 \left( \bar{\lambda}_{\vec{N}^+} - \xi \bar{\lambda}_{\Delta \vec{S} \vec{N}^+} \right)^2 \right. \\
   \left. + \sum_{\vec{N},\vec{N'}} \left| \tilde{f}^{(1)}_{\vec{N}} \right|^2 e^{-  \bar{\lambda}_{\vec{N}'} t }  \left( \bar{\lambda}_{\vec{N}'} - \xi \bar{\lambda}_{\vec{N}} \right) \left( -\delta_{\bar{\lambda}_{\vec{N}'} , \bar{\lambda}_{\vec{N} - \vec{N}'}}  + \delta_{\bar{\lambda}_{\vec{N}'} \neq \bar{\lambda}_{\vec{N} - \vec{N}'}} \frac{\bar{\lambda}_{\vec{N}'} - \xi \bar{\lambda}_{\vec{N}}}{\bar{\lambda}_{\vec{N} - \vec{N}'}  - \bar{\lambda}_{\vec{N}'}} \right) \right],
\end{multline}
\begin{multline}
K^{(2)}_\mathrm{f}(t) = t \, \frac{4(2 \pi)^4}{ (L_1 L_2)^2} \left[ \frac{t}{2} \sum_{\vec{N}^+}  e^{-\bar{\lambda}_{\vec{N}^+} t } \sum_{\mathclap{\vec{S},\vec{S}', \vec{S}' \ne \vec{S}}} \left| \tilde{f}^{(1)}_{\Delta \vec{S} \vec{N}^+} \right|^2
 \left( \left( \bar{\lambda}_{\vec{N}^+} - \frac{1}{4} \bar{\lambda}_{\Delta \vec{S} \vec{N}^+} \right)^2 - D_{\vec{S} \vec{N}^+, \vec{S}'\vec{N}^+}^2  \right) \right. \\
 		\left.  + \sum_{\vec{N},\vec{N}'} \left| \tilde{f}^{(1)}_{\vec{N}} \right|^2 e^{-\bar{\lambda}_{\vec{N}'} t } \left( \frac{3}{16} \bar{\lambda}_{\vec{N}} - \bar{\lambda}_{\vec{N}'} + \delta_{\bar{\lambda}_{\vec{N}'} \neq \bar{\lambda}_{\vec{N} - \vec{N}'}} \frac{\left(\bar{\lambda}_{\vec{N}'} - \bar{\lambda}_{\vec{N}}/4 + D_{\vec{N}, \vec{N}'}  \right)\left( \bar{\lambda}_{\vec{N} - \vec{N}'} - \bar{\lambda}_{\vec{N}}/4 - D_{\vec{N}, \vec{N}'}  \right)}{\bar{\lambda}_{\vec{N}-\vec{N}'} - \bar{\lambda}_{\vec{N}'}} \right) \right],
\end{multline}
\end{subequations}
\end{widetext}
where we defined~$\vec{S} \vec{N}^+ \equiv \{s_1 n_1^+,s_2 n_2^+\}$,~$\Delta \vec{S} \vec{N}^+ \equiv (\vec{S} - \vec{S}')\vec{N}^+ $,~$D_{\vec{N}, \vec{N'}} = i(2\pi)^2 \left(n_1 n_2' - n_1' n_2 \right)/(2L_1 L_2)$, and~$\delta_{\bar{\lambda}_{\vec{N}'} \neq \bar{\lambda}_{\vec{N} - \vec{N}'}} = 1$ if~$\bar{\lambda}_{\vec{N}'} \neq \bar{\lambda}_{\vec{N} - \vec{N}'}$ and~0 otherwise.  Note that the precise form of the matrices $c^{(\vec{N}^+)}_{\vec{S}\vec{S}'}$ does not matter since they cancel out in traces, but the presence of the sums over degenerate subspaces in the first terms in the above expressions is an artifact of needing to treat the degenerate subspaces properly.

We may now take the limit~$L \to \infty$.  The first term in each expression above vanishes in this limit (essentially because~$L^{-4} \sum_{\vec{N}} \to L^{-2} \int d^2 k \to 0$); for the same reason, the terms containing~$\delta_{\bar{\lambda}_{\vec{N}'} , \bar{\lambda}_{\vec{N} - \vec{N}'}}$ also vanish.  We then obtain equation~\eqref{eq:heatkernelexpr} with
\begin{subequations}
\be
I_\mathrm{s}(k^2 t) = \frac{4}{k^4} \, \Pcal \int d^2 q \,  e^{- q^2 t }  \, \frac{ ( q^2 - \xi k^2 )^2 }{ k^2 - 2 {\vec{q}} \cdot \vec{k}},
\ee
\begin{multline}
I_\mathrm{f}(k^2 t) = \frac{4}{k^4} \, \Pcal \int d^2 q \,  e^{- q^2 t }  \left(-\frac{1}{16} k^2 \right. \\ \left. + \frac{ ( q^2 - k^2/4)^2 + (\vec{k} \times \vec{q})^2/4}{ k^2 - 2 {\vec{q}} \cdot \vec{k}}\right),
\end{multline}
\end{subequations}
where~$q = |\vec{q}|$,~$k = |\vec{k}|$,~$\vec{k} \times \vec{q} = k_1 q_2 - k_2 q_1$, and~$\Pcal$ denotes a Cauchy principal value (which comes about since terms in which the denominator vanishes are excluded in the discrete sums).  After integration, we obtain~\eqref{eq:scalar} and~\eqref{eq:dirac}.


\end{document}